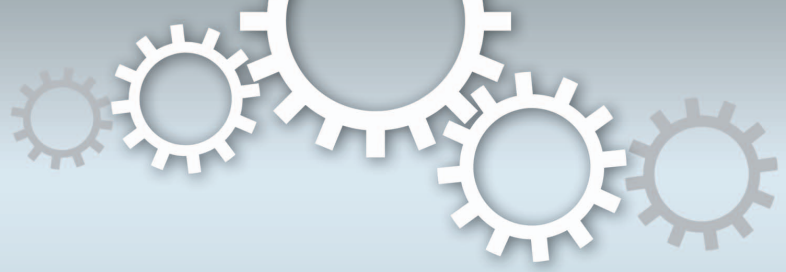

# Tunable band gap in few-layer graphene by surface adsorption

Ruge Quhe[1,2]*, Jianhua Ma[3]*, Zesheng Zeng[1], Kechao Tang[1], Jiaxin Zheng[1,2], Yangyang Wang[1], Zeyuan Ni[1], Lu Wang[4], Zhengxiang Gao[1], Junjie Shi[1] & Jing Lu[1]

[1]State Key Laboratory of Mesoscopic Physics and Department of Physics, Peking University, Beijing 100871, P. R. China, [2]Academy for Advanced Interdisciplinary Studies, Peking University, Beijing 100871, P. R. China, [3]School of Physics and Nuclear Energy Engineering, Beihang University, Beijing 100191, P. R. China, [4]Department of Physics, University of Nebraska at Omaha, Omaha, Nebraska 68182-0266.

There is a tunable band gap in ABC-stacked few-layer graphene (FLG) via applying a vertical electric field, but the operation of FLG-based field effect transistor (FET) requires two gates to create a band gap and tune channel's conductance individually. Using first principle calculations, we propose an alternative scheme to open a band gap in ABC-stacked FLG namely via single-side adsorption. The band gap is generally proportional to the charge transfer density. The capability to open a band gap of metal adsorption decreases in this order: K/Al > Cu/Ag/Au > Pt. Moreover, we find that even the band gap of ABA-stacked FLG can be opened if the bond symmetry is broken. Finally, a single-gated FET based on Cu-adsorbed ABC-stacked trilayer graphene is simulated. A clear transmission gap is observed, which is comparable with the band gap. This renders metal-adsorbed FLG a promising channel in a single-gated FET device.

Opening a tunable band gap in graphene without degrading its high carrier mobility is crucial for its utilization in nanoelectronics[1]. So far, there are two feasible schemes that can reach this goal for bilayer graphene. One is the application of an external electric field to destroy the inversion symmetry of bilayer graphene[2–10]. Both theoretical[6,7,9,10] and experimental[2] research demonstrated that a variable band gap up to 0.25 eV can be engineered in bilayer graphene (BLG) by applying a perpendicular electric field to break the inversion symmetry of BLG. This band gap enables the fabrication of an effective bilayer graphene FET with the current on/off ratio of 100 and 2000 at room temperature and 20 K, respectively[11]. The main drawback of this scheme is the requirement of two individual gates to control the band gaps and vary the charge carrier concentration simultaneously. From a technical point of view, it is desirable to control a transistor's conductance by only one gate. An alternative scheme is using single-side adsorption to break the inversion symmetry of BLG. Single-side adsorption with metal atoms such as potassium[12,13] and aluminum[14] or molecules such as water[15], oxgen[16], benzyl viologen[16], tetracyanoquinodimethane (TCNQ)[17,18], tetrafluoro-tetracyanoquinodimethan (F4-TCNQ)[19], decamethylcobaltocene[20], and 3,6-difluoro-2,5,7,7,8,8-hexacyano-quinodimethane (F2-HCNQ)[20] is found to open a band gap of BLG, and this band gap is tunable by changing the adsorbate coverage. Based on the surface adsorption scheme, single-gated BLG-based FETs have been fabricated[13,15,16,19], with an improved current on/off ratio up to 45 at room temperature[14].

With the progress in BLG, more and more attention is turned to few-layer graphene (FLG). Theoretical study showed that a vertical electric field can also open a tunable band gap comparable to bilayer graphene in ABC-stacked FLG[21–23] but fail in ABA-stacked FLG[21,23]. Such an electrically tunable band gap has been recently observed[24–30]. The natural next step should be exploration of the possibility whether adsorbate doping is able to open a band gap in FLG and fabrication of a single-gated FLG-based FET if it can. In this article, we present the first systematic theoretical investigation of the effects of single-side adsorption of metal atoms (including K, Al, Cu, Au, Ag, and Pt) on the electronic structure of ABC-stacked FLG. A band gap is opened in the most cases as a result of the break of inversion symmetry in ABC-stacked FLG, and its size is tunable by changing the dopant concentration and species. Surprisingly, even the band gap of ABA-stacked FLG can be opened if the bond symmetry is broken by surface adsorption. Finally, we simulate a single-gated Cu-adsorbed ABC-stacked trilayer FET by using *ab initio* quantum transport calculation and verify the existence of a transport gap in this device. Therefore, FLG with surface adsorption can serve as the channel of a single-gated FET.

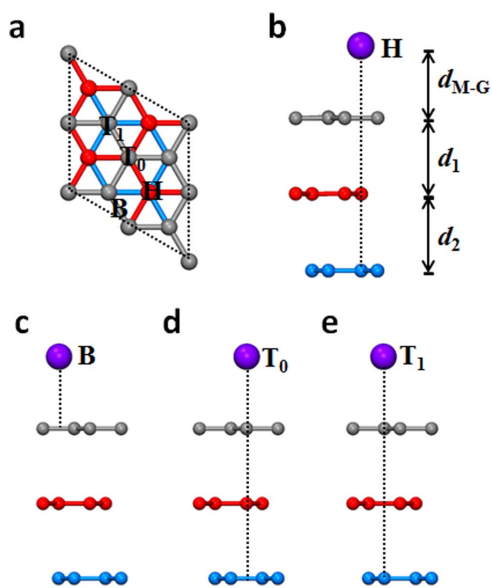

**Figure 1 | Possible adsorption sites of the metal atom on FLG.** (a) Lattice structure of ABC-stacked graphene trilayer; gray/red/blue indicate links on the top/middle/bottom layers, and $H/B/T_0/T_1$ distinguish the four adsorption sites for a metal adatom on graphene: hollow center ($H$), bridge center ($B$), and two top sites ($T_0$ and $T_1$) in a $2 \times 2$ graphene supercell. (b–d) Side views of the metal-adsorbed FLG with metal atom at the (b) $H$ site, (c) $B$ site, (d) $T_0$ site, and (e) $T_1$ site. The $T_0$ site is right above the carbon atoms of the top and middle layers, and the hollow center of the bottom layer, while the $T_1$ site is above the carbon atoms of the top and bottom layers, and the hollow center of the middle layer.

## Results

**Metal-adsorbed ABC-stacked trilayer graphene.** As illustrated in Fig. 1, three possible adsorption sites of a single metal atom on trilayer graphene are considered: hollow ($H$), top ($T$) and bridge ($B$). Notice that there are actually two types of $T$ site: one (marked as $T_0$) with the metal atom above not only a carbon atom of the uppermost graphene but also a carbon atom of the second graphene layer; the other (marked as $T_1$) with the metal atom located above a carbon atom in the uppermost graphene layer and a hexagon center in the second graphene layer. Nearly irrespective of the coverage, the two simple metals (K and Al) are found to favor the $H$ site while the four noble metals (Au, Ag, Cu and Pt) favor the $T_0$ site in FLG. This is consistent with the experimental observation of metal on FLG: alkali metal Cs favors the $H$ site and noble metal Au favors the $T$ site[31,32]. Interestingly, while K, Al, Au, and Ag atoms adsorbed on (single layer graphene) SLG favor the same sites as on FLG, Cu and Pt atoms favor the $B$ site on single layer graphene (SLG) instead of the $T$ site according to previous theoretical results[33].

The metal adsorption causes a slight buckling of the uppermost graphene layer, with values ranging from 0.01 to 0.06 Å. The shape of middle and bottom graphene layers and the two interlayer distances $d_1$ and $d_2$ (with an increase of less than 0.01 Å) are nearly intact. Fig. 2a shows the dependence of the distance between the metal atom and the uppermost graphene sheet ($d_{M-G}$) on the coverage. The $d_{M-G}$ of all kinds of metal atom (2.1 ~ 3.5 Å) shows an increasing tendency with the increasing coverage. This monotonic behavior is due to the fact that the Mulliken charge per adatom $Q_M$ transferred from metal to graphene decreases with the increasing coverage (see Fig. 2b), which leads to a weaker Coulomb attraction between the metal atom and graphene. The $d_{M-G}$ value of a given coverage for each kind of metal adatom increases with the metal atomic radius ($r$), namely $d_{M-G}$ (K) > $d_{M-G}$ (Al) and $d_{M-G}$ (Au) > $d_{M-G}$ (Ag) > $d_{M-G}$ (Pt) > $d_{M-G}$ (Cu). According to the size of $d_{M-G}$, the graphene-metal bonds can be classified into two types: Those with $d_{M-G} > 2.7$ Å (Au, Ag, K over the checked coverages, Al with $\theta = 1/3$, and Cu with $\theta < 1$) is an ionic bond, which tends to dope graphene, while those with $d_{M-G} < 2.7$ Å (Pt, Al with $\theta < 1/3$, and Cu with $\theta < 1$) is a covalent bond, which tends to destroy graphene band structure. Subsequent band structure calculations will verify this classification.

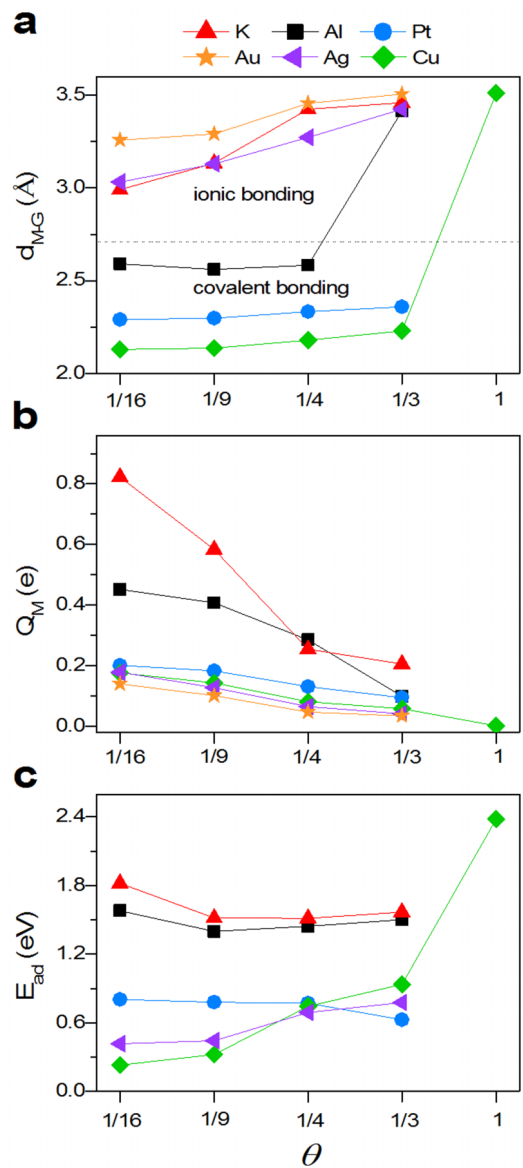

**Figure 2 |** (a) Distance between the metal adatom and uppermost graphene layer, (b) Mulliken charge transfer from a metal adatom to graphene, and (c) adsorption energy of the metal adatom on ABC-stacked trilayer graphene as a function of the coverage. We only consider the case of Cu for the coverage of $\theta = 1$, since the size of other metals is too large to fill the $1 \times 1$ lattice. The adsorption energies of Au and Ag at the same coverage nearly coincide.

All types of metal adsorption tend to donate less charge to trilayer graphene with the increasing $\theta$, and the charge transfer per adatom $Q_M$ of the four noble metal atoms to trilayer graphene is generally smaller than that of the two simple metals. The $Q_M$ of the two simple metals decreases sharp when the $\theta$ increases (K: from 0.9 to 0.1 $e$; Al: from 0.5 to 0.1 $e$) (Fig. 2b). By contrast, the change of $Q_M$ of the noble metals is rather slow with a narrow magnitude of 0 ~ 0.2 $e$. The adsorption energy of one metal atom on FLG is defined as

$$E_{ad} = E_G + E_M - E_{M/G}, \qquad (1)$$

<mark>where $E_G$, $E_M$ and $E_{M/G}$ are the relaxed energy of FLG, the isolated metal atom, and the combined system, respectively. As shown in Fig. 2c, the two simple metals are adsorbed more strongly ($E_{ad} > 1.1$ eV) on trilayer graphene than the four noble metals ($E_{ad} < 0.8$ eV except for the case of Cu at $\theta = 1$), due to the generally larger charge transfer from the two simple metals to trilayer graphene. With the increasing coverage, $E_{ad}$'s of the two simple metals decrease firstly and then increase, $E_{ad}$'s of Au, Ag, and Cu always increase, but $E_{ad}$ of Pt always decreases. The change of $E_{ad}$ with the coverage depends on the competition between the decreasing metal-graphene and the increasing metal-metal interactions with the increasing coverage. The decreasing metal-graphene interaction with the increasing coverage is attributed to the reduction of $Q_M$ (Fig. 2b), while the increasing metal-metal interaction with the increasing coverage is attributed to the gradual formation of a metallic bond. The exceptionally large $E_{ad}$ of 2.8 eV of Cu on trilayer graphene with $\theta = 1$ indicates that strong metallic bonds have formed between the Cu atoms.

The band structures of ABC-stacked trilayer graphene adsorbed by different metal atoms at the given coverage of $\theta = 1/4$ are provided in Fig. 3. Strong band hybridization takes place between graphene and adsorbed Al, Pt, and Cu atoms with $d_{M-G} < 2.7$ Å, confirming the formation of a covalent bond. By contrast, the graphene bands are almost intact for Au, Ag, and K adsorption with $d_{M-G} > 2.7$ Å, indicative of an ionic bonding. All the metals open a band gap near the $K$ point of graphene, forming a Mexican hat structure. The band gap is 0.237, 0.000, 0.185, 0.201, 0.190 and 0.082 eV for K, Al, Cu, Ag, Au, and Pt adsorption, respectively. The closing of the band gap near the Dirac point for Al adsorption is due to the large band distortion induced by the strong hybridization between graphene valence band and the Al 3p-derived bands (See Fig. 3b), which pushes the valence band maximum (VBM) state of graphene at the $K$ point upwards. The relatively smaller band gap by Pt adsorption is also ascribed to the rise of VBM state of graphene at the $K$ point caused by the hybridization with the Pt 5d/6s-derived bands (Fig. 3d). The Pt 5d and 6s bands are hardly separable, due to the close energy level between 5d and 6s orbitals in Pt atoms. Apparently, K has the strongest capability to open the band gap of trilayer graphene, followed by Cu/Ag/Au, and Pt, and Al is the weakest one at this coverage. The opened band gaps by K/Cu/Ag/Au adsorption approach the theoretical maximum band gap opened by a uniform vertical electric field (0.226 eV)[23]. Therefore, the ability to open a band gap by adsorption of the four metals is comparable with that of the electric field. In the most cases, the band gaps are direct, similar to the case under a vertical electric field[23]. The band structures in Fig. 3 also confirm that trilayer graphene is apparently n-doped by K, Al, and Cu adsorption, with $E_f - E_D = 0.31$, 0.17 and 0.21 eV, respectively. The bands derived from the metal outmost s/p valence band cross the Fermi level ($E_f$) except for Pt.

Fig. 4 shows the band structures of Cu-adsorbed trilayer graphene at four coverages. The Dirac point of trilayer graphene is folded to the $\Gamma$ point due to Brillouin Zone folding when $\theta = 1/9$ and $1/3$ (Figs. 4b and 4c). A direct band gap is opened near the Dirac point with values of $\Delta = 0.152$, 0.177, 0.182 and 0.073 eV for $\theta = 1/16$, 1/9, 1/3 and 1, respectively. The band gaps as a function of the coverage for the six metals are shown in Fig. 5a. The opened band gaps generally increase with the increasing coverage, except for Al adsorption with a minimum at $\theta = 1/4$ and Cu adsorption with a maximum at $\theta = 1/3$. This band gap opening can be attributed to the built-in of electric field vertical to the graphene sheet due to the metal-graphene charge transfer (Fig. 5b), which breaks the inversion symmetry of trilayer graphene. In terms of a simplified parallel plate capacitor model, the potential difference between the third graphene layer and the first graphene layer can be written as $U_3 - U_1 = -\alpha(\Delta n_2 + 2\Delta n_3) \neq 0$, where $\alpha = ed_0/\varepsilon_0\kappa$, $\varepsilon_0$ is the vacuum permittivity, $\kappa$ the dielectric constant, and $\Delta n_i$ is the electron density change of the $i$ layer. We have assumed $d_{M-G} = d_1 = d_2 = d_0$. With the increasing coverage, the charge provided by per metal atom ($Q_M$) generally decreases but the charge transfer averaged over the surface C atom number $Q_C$ (namely, charge transfer density) generally increases for K, Au, and Pt adsorption, and peaks at $\theta = 1/4$ for Al, Cu, and Ag adsorption, as shown in Fig. 5c. Generally speaking, a larger $Q_C$ causes a stronger perpendicular built-in electric field of graphene and thus a lager band gap $\Delta$. In fact, given a metal, the band gap $\Delta$ and the charge transfer density $Q_c$ have an approximate linear relation, as shown in Fig. 5d. The exceptional enhancement in the band gap for Al and Ag adsorption at $\theta = 1/3$ is attributed to existence of an extra mechanism of band gap opening, namely, the breaking of bond symmetry, at this coverage beside the well-known mechanism of the breaking of the inversion symmetry. The band gap opening via the breaking of bond symmetry has already been found in Li-adsorbed SLG[34] and alkali atom-adsorbed single layer silicene[35].

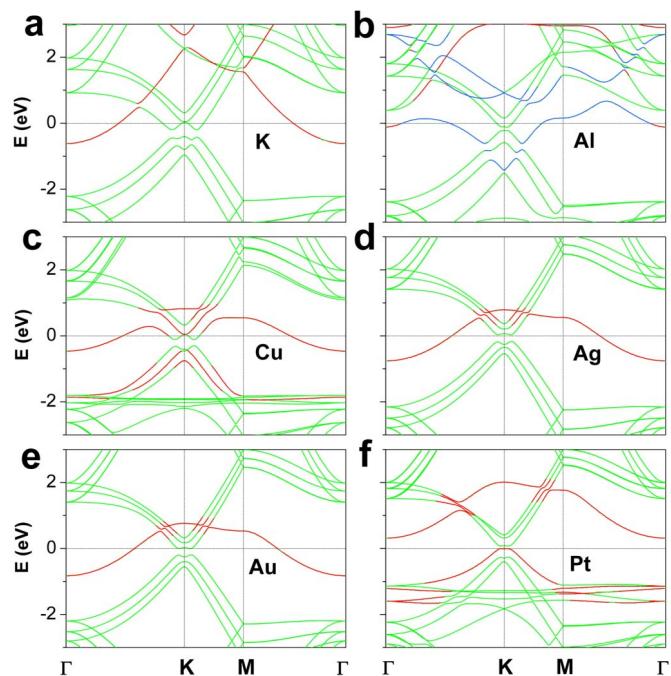

**Figure 3 | Band structures of (a) K, (b) Al, (c) Cu, (d) Ag, (e) Au, and (f) Pt-adsorbed ABC-stacked trilayer graphene with a coverage of $\theta = 1/4$.** The red denotes the bands with weight projected on the metal outmost $s$ valence band in (a–e) and on both the 5d and 6s bands in (f), and the blue denotes the bands with weight projected on the Al outmost 3p valence band. The Fermi level is set to zero.

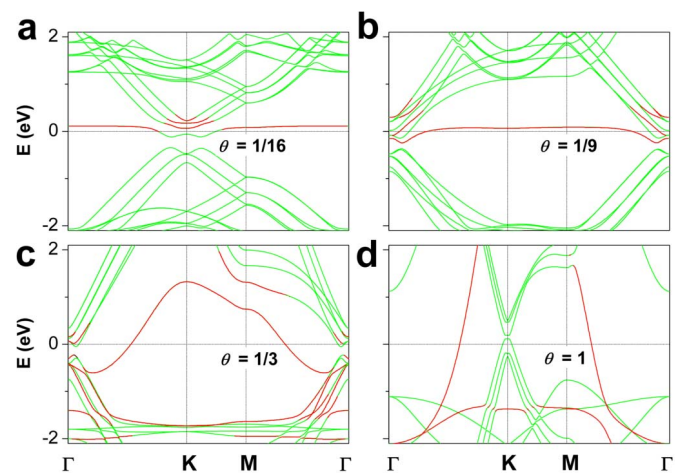

**Figure 4 | Band structures of Cu-adsorbed ABC-stacked trilayer graphene with the coverage of $\theta =$ (a) 1/16, (b) 1/9, and (c) 1/3, and (d) 1.** The red denotes the bands with weight projected on the metal outmost $s$ valence band. The Dirac point of graphene in (b) and (c) is folded to the $\Gamma$ point due to the reduction of the first Brillouin zone. The Fermi level is set to zero.

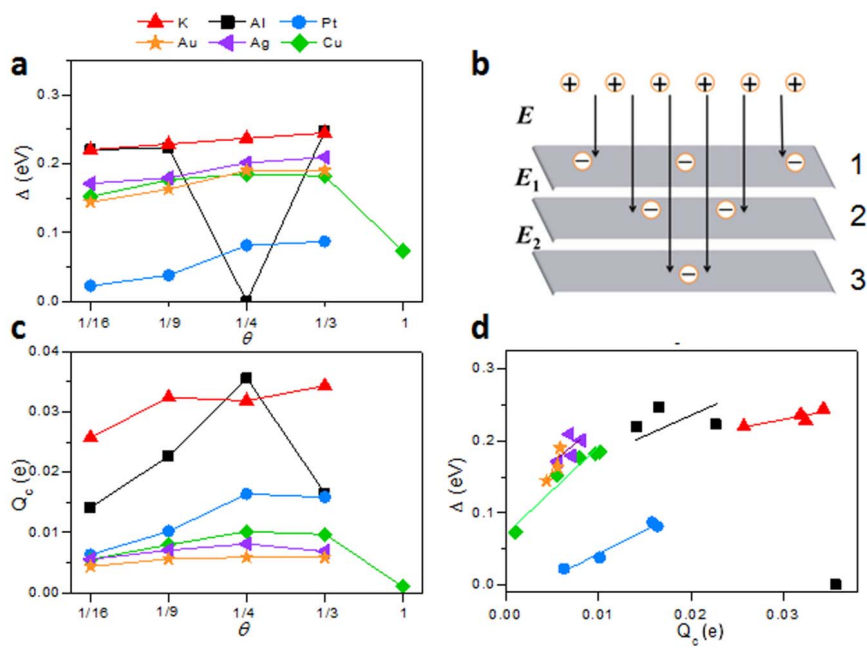

**Figure 5** | (a) Band gaps of graphene in metal-adsorbed ABC-stacked trilayer graphene as a function of the adsorption coverage. (b) Schematic illustration of equivalent vertical electric field induced by metal single-side adsorption. $E$ represents the electric field between metal and graphene and $E_1$ and $E_2$ represent the electric field between the graphene layers. (c) Charge transfer density in metal-adsorbed ABC-stacked trilayer graphene as a function of the adsorption coverage. (d) Band gaps of graphene in metal-adsorbed ABC-stacked trilayer graphene as a function of the charge transfer density. There is an approximately linear relation between the band gap and the charge transfer density.

In the wide coverage range, the adsorbed metals can generally be divided into three groups in terms of the size of the opened band gap. Group I (K/Al) opens the largest band gap with value ranging from 0.220 to 0.248 eV under different coverages (except for the abnormal band gap induced by Al adsorption at $\theta = 1/4$), followed by group II (Cu/Ag/Au) with band gaps of 0.073 ~ 0.209 eV, and group III (Pt) has the least band gap with value of 0.023 ~ 0.087 eV. Given a coverage, the largest band gaps opened by K and Al adsorption are attributed to their largest charge transfer density $Q_C$. Although K donates more electron than Al at a given coverage (Fig. 2b), it has a larger metal-graphene distance (Fig. 2a). The two factors lead to a similar band gap for K and Al adsorptions. Given a coverage, the band gaps are similar for Cu, Ag, and Au adsorptions due to the similar $Q_C$. As stated previously, the exceptionally smaller band gap by Pt adsorption is attributed to the rise of the graphene valence band pushed by the Pt $5d/6s$-derived bands below $E_f$.

**Metal-adsorbed ABC-stacked thicker FLG.** We investigate K adsorption on ABC-stacked tetralayer and pentalayer graphene and show the electronic structures in Figs. S1 and S2. A direct band gap is opened in K-adsorbed tetralayer graphene at $\theta = 1/9$ and 1/4 with values of 0.054 and 0.117 eV, respectively. No band gap is opened in K-adsorbed tetralayer graphene at $\theta = 1/3$ and pentalayer graphene at $\theta = 1/4$ due to a stronger band hybridization between graphene and metal atom. These band gap values in thicker FLG are apparently smaller than those in ABC-stacked trilayer ($\Delta = 0.228$, 0.237 and 0.245 for $\theta = 1/9$, 1/4 and 1/3, respectively) and bilayer ($\Delta = 0.264$, 0.274 and 0.280 eV for $\theta = 1/9$, 1/4 and 1/3, respectively) graphene at the same coverage. Therefore, the band gap decreases with the increasing layer for K adsorption. From a technical point of view, thicker K-adsorbed FLG is unfavorable for the application as the channel of FET device.

**Metal-adsorbed ABA-stacked trilayer graphene.** Compared with ABC-stacked FLG, it is much difficult to open a band gap in ABA-stacked FLG because it has no inversion symmetry. According to the previous theoretical and experimental results[23,25], a vertical electric field fails to open a band gap in ABA-stacked FLG graphene but increases the overlap between the conduction and valence bands instead[24]. However, a band gap of 0.054 eV is opened in Al-adsorbed ABA-stacked trilayer graphene at $\theta = 1/3$, as shown in Fig. 6a. The band gap opening in these cases purely originates from the bond symmetry breaking[34,35] rather than the inversion symmetry. The corresponding space charge distribution is provided in Fig. 6b, illustrating the unsymmetrical charge distribution among different C-C bonds.

**Transport properties of ABC-stacked trilayer graphene.** We simulate the transport properties of ABC-stacked trilayer graphene before and after Cu atoms adsorption. The two-probe model is presented in Fig. 7a, with the semi-infinite pure ABC-stacked trilayer graphene itself as electrodes. The channel length is $L_{ch} = $ 10 nm, with Cu adsorption coverage of $\theta = 1$. We show the zero-bias transmission spectrum of the device with Cu adsorption in Fig. 7b, where a gap of 0.080 eV is observed. It originates from a gap of the same size in the projected density of states (PDOS) of the channel graphene, as show in Fig. 7c, because the transmission coefficient of the device, $T(E)$, is connected with the PDOS of the channel and the two electrodes via the relation:

$$T(E) \propto \frac{g_{ch}(E)g_L(E)g_R(E)}{g_{ch}(E)g_L(E) + g_{ch}(E)g_R(E) + g_L(E)g_R(E)} \quad (2)$$

where $g_{ch}(E)$ and $g_{L/R}(E)$ are the PDOS of the channel and the left/right lead, respectively. Both gaps are can be attributed to the band gap of $\Delta = 0.082$ eV of the corresponding infinite Cu-adsorbed ABC-stacked trilayer graphene. By contrast, there is no gap for pure trilayer graphene device (Inset in Fig. 7b).

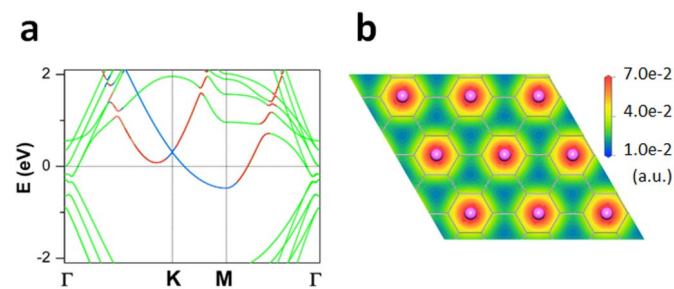

**Figure 6** | (a) Band structure and (b) electron distribution on the uppermost layer graphene of Al-adsorbed ABA-stacked trilayer graphene with the coverage of $\theta = 1/3$. The blue and red lines denote the bands with weight projected on the Al $3p$ and $4s$-derived bands, respectively. The Dirac point of graphene is folded to the $\Gamma$ point due to the reduction of the first Brillouin zone. The Fermi level is set to zero.

The appearance of the transport gap upon Cu adsorption is also reflected from a change of the transmission eigenchannel at $E - E_f = -0.03$ eV and at the (1/3, 0) point of the $k$-space. As displayed in Fig. 7d, the transmission eigenvalue at the point is 0.94 for pure trilayer graphene, and the incoming wave function is scattered little and most of the incoming wave is able to reach to the other lead. By contrast, the transmission eigenvalue at this point nearly vanishes upon Cu adsorption, and the incoming wave function is nearly completely scattered and unable to reach to the other lead. The carrier mobility of FLG is expected not be degraded significantly by metal adsorption since that of bilayer graphene is not significantly degraded by Al and molecule single-side adsorption[14,19].

## Discussion

It is interesting to make a comparison between metal-adsorbed trilayer graphene and metal-adsorbed bilayer graphene in view of the fact the band gap opening has been successfully observed in the latter. The metal-graphene distance, adsorption energy, charge transfer, and band gap of metal-adsorbed bilayer graphene are provided in Figs. S3 and S4. In all the checked adsorptions, a band gap (it is also direct in the most cases) is opened. Similar to its trilayer counterpart, group I (K/Al) opens the largest band gap (0.231 ~ 0.280 eV) in bilayer graphene, followed by group II (Cu/Ag/Au) with band gaps of 0.057 ~ 0.267 eV, and group III (Pt) has the least band gap (0.016 ~ 0.187 eV). Given a metal, the band gap and the charge transfer density $Q_c$ also have an approximate linear relation. In Cu-adsorbed bilayer graphene, the band gaps opened at $\theta = 1/9$ and 1/3 are apparently larger than those at other coverages. The enhancement of the band gap at $\theta = 1/9$ and 1/3 is also attributed to the breaking of bond symmetry (See more in Fig. S5)[34]. The band structures of metal-adsorbed bilayer graphene at a coverage of $\theta = 1/4$ are provided in Fig. S6.

Compared with bilayer graphene case, the distances between the metal atom and the uppermost graphene of the two simple metals $d_{M-G}$ are increased by about 0.3 ~ 0.6 Å while those of the four noble metals are reduced slightly by 0 ~ 0.04 Å under a given coverage in ABC-stacked trilayer graphene case (Fig. 8a). As shown in Fig. 8b, the transferred charge per adatom $Q_M$ to trilayer graphene is almost identical to that transferred to the bilayer graphene, except for Al adsorption at $\theta = 1/16$ and 1/9, where the $Q_M$ differs by about 0.1$e$ between trilayer and bilayer graphene.

The difference of the adsorption energy between metal-adsorbed trilayer and bilayer graphene is provided in Fig. 8c. At $\theta = 1/16$, all the metal atoms favor adsorption on trilayer graphene. With the increasing $\theta$, adsorption on bilayer graphene is gradually preferred. Noticeably, the adsorptions of Au and Ag on trilayer graphene are stronger than or equal to those on bilayer graphene under all considered coverages. The stronger adsorption of Au and Ag on the former can be attributed to the reduced metal-graphene distance. Experimentally, the increase of the layer number indeed strengthens the adsorption of Au atom on graphene[36]. The adsorptions of K and Al on trilayer graphene are weaker than on bilayer graphene at high coverage (K: $\theta > 1/9$; Al: $\theta > 1/4$) as a result of the increased metal-graphene distance $d_{M-G}$ in trilayer graphene and the small change in the charge transfer $Q_M$. Although the $d_{M-G}$ value of K and Al is larger in trilayer graphene at low coverage too, the $Q_M$ value is also larger, and the latter factor dominates the adsorption energy. As a result, the K and Al atom are adsorbed more strongly on trilayer graphene than the bilayer one at $\theta < 1/9$ and $\theta < 1/4$ respectively.

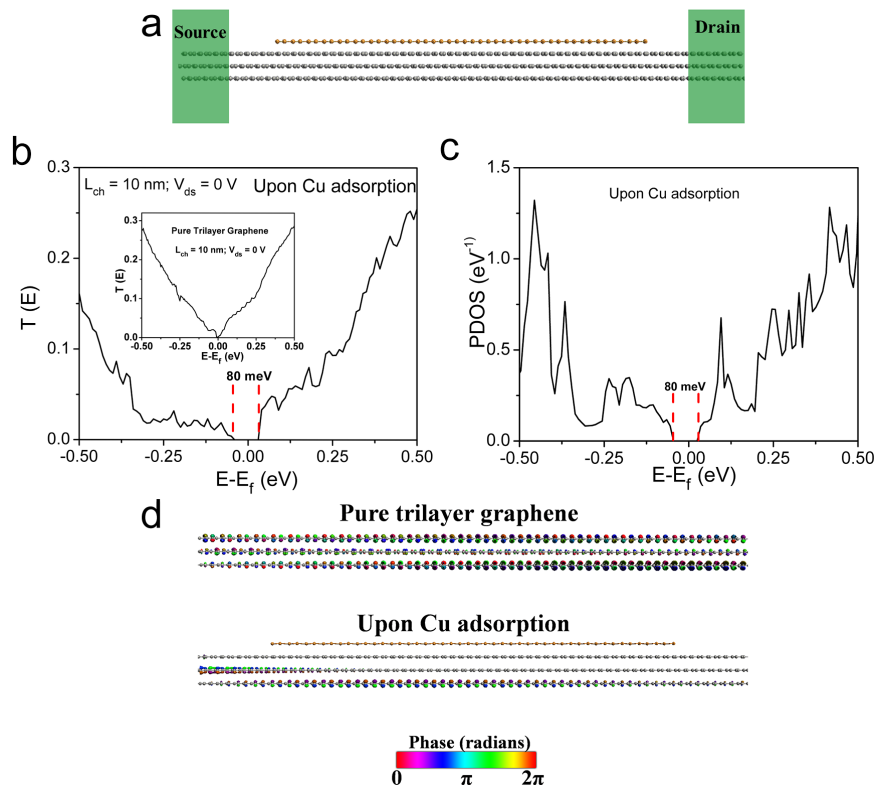

**Figure 7** | (a) Two-probe model of ABC-stacked trilayer graphene adsorbed by Cu atoms. The length of the channel $L_{ch}$ is $L_{ch} = 10$ nm. Gray ball: C; yellow ball: Cu (b) Transmission spectrum of this model under zero bias. Inset: transmission spectrum of a pure ABC-stacked trilayer graphene with the same $L_{ch}$ under zero bias. (c) Projected density of states (PDOS) of the channel trilayer graphene. (d) Transmission eigenstates of a trilayer graphene before and after Cu atoms adsorption at $E - E_f = -0.03$ eV and $k = (1/3, 0)$. The isovalue is 0.4 a.u.

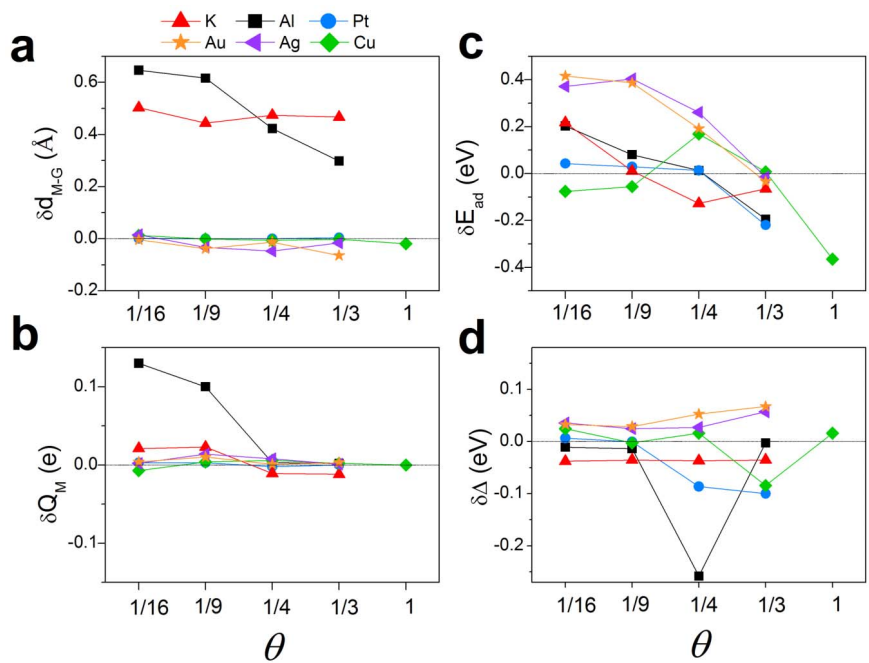

**Figure 8** | Differences in the (a) metal-graphene distance, (b) Mulliken charge transfer from a metal adatom to graphene, (c) adsorption energy, and (d) band gap between ABC-stacked trilayer and bilayer graphene adsorbed by different metals. The dotted lines are guides to the eye.

As shown in Fig. 8d, the band gaps of K/Al-adsorbed trilayer graphene are smaller by 0.01 ~ 0.04 eV than those of their bilayer counterparts at a given coverage. On the contrary, the band gaps opened in trilayer graphene by Ag/Au adsorption are about 0.02 ~ 0.07 eV larger than those in their bilayer counterparts. This difference can also be understood by a difference in the equivalent vertical electric field ($E_\perp$), which chiefly depends on the difference in $d_{M-G}$ here since the difference in $Q_M$ and $Q_C$ is quite small in the most cases. The $d_{M-G}$ values of K/Al-adsorbed trilayer graphene are apparently larger than those of their bilayer counterpart, and $E_\perp$'s in trilayer graphene are thus weaker, resulting in the smaller band gaps. The especially larger $Q_M$ in Al-adsorbed trilayer at $\theta = 1/16$ and 1/9 suppresses the band gap difference. By contrast, the $d_{M-G}$ values of Ag/Au-adsorbed trilayer graphene are smaller than those of their bilayer counterpart, and $E_\perp$'s are thus stronger, leading to the larger band gaps. The difference between the band gap of Cu- and Pt-adsorbed trilayer and bilayer graphene shows an oscillation behavior around the zero reference line, since their $d_{M-G}$ values between trilayer and bilayer graphene in the two cases are too small to cause the positive- or negative-definite band gap difference.

Experimentally, clustering of metal atoms such as Fe, Au, and Pt on graphene surface has been frequently observed[31,36–40]. To check the effect of atomic clustering on the band gap of FLG, we further carry out a calculation based on the structure of 4-atom K cluster on a ABC-stacked $4 \times 4$ supercell of trilayer graphene (the corresponding coverage is thus $\theta = 1/4$). As shown in Fig. S7a, the four K atoms form a tetrahedron with K-K bond length of 4.37 or 5.18 Å. Its total energy is 0.040 eV per K atom lower than that of individual K atom superlattice at the same coverage. Each of the bottom K atoms donate $Q_M = 0.182$ $e$ to the graphene while the top K atom donate only 0.040 $e$, showing a weaker $n$-dope ability than the individual K atom of the superlattice with $Q_M = 0.255$ $e$. In Fig. S7b, we can see that there is still a direct band gap of 0.231 eV around the Dirac cone. Compared with the case of individual K atom superlattice at the same coverage, the band gap in ABC-stacked trilayer graphene introduced by the K cluster is just 0.013 eV smaller. Therefore, the band gap in FLG appears not significantly degraded even if the adsorbed metal atoms form clusters.

The opened band gaps of FLG are often above or below $E_f$. Experimentally, Al film adsorption on BLG causes a band gap opening and heavy $n$-type doping of BLG[14], a result in agreement with our calculations (Fig. S2). This band gap is pushed back to $E_f$ by a back gate voltage of −40 V. Our recent calculations show that alkali metal single-side adsorption on silicene also causes a band gap opening and heavy $n$-type doping, and the band gap is pushed back to $E_f$ by a gate voltage of about −30 V[35]. Note that in our checked metals, alkali metal and Al have the highest electron doping ability. The band gap opened in FLG by metal adsorption should be able to be pushed back to $E_f$ by an experimentally accessible gate voltage.

In summary, we reveal for the first time that single-side adsorption of metal atom is able to effectively open a band gap in both ABC- and ABA-stacked FLG. Therefore, the FET device with metal-adsorbed FLG as the channel only requires one gate. The band gap is generally proportional to the charge transfer density. The ability to open a band gap decreases in this order of K/Al > Cu/Ag/Au > Pt. Trilayer graphene with surface dopant is a promising channel, competitive with BLG with surface dopant, in a single-gated graphene FET device.

## Methods

Two dimensional structures of ABC- and ABA-stacked FLG with metal adsorption are simulated within a hexagonal supercell. To feature different metal adsorption coverage ($\theta$) and study its possible influence on the band gap, we create supercell out of an $m \times m$ ($m = 1, \sqrt{3}, 2, 3,$ and 4) graphene primitive cell, and place one metal adatom above the top graphene layer within the supercell. We define the coverage $\theta$ as the ratio of the number between the metal atom and primitive graphene cell per supercell, and one has $\theta = 1/m^2$. The in-plane lattice constant of graphene is taken from that of experimental value, $a = 2.46$ Å[41]. A vacuum buffer space of at least 20 Å is set to prevent the interaction between adjacent slabs. Recently, well-ordered superlattice of Cs atom single layer on FLG has been observed by using scanning tunneling microscopy (STM) measurements as a result of strong repulsion between negatively charged Cs atom[32]. Individual Au atoms have also been clearly observed on clear FLG in high angle annular dark field (HAADF) images, although no individual Au atoms are observed on clear SLG due to a stronger interaction between Au and FLG than that between Au and SLG[31]. These experimental results give a support to our ordered metal atom single layer superlattice model on FLG.

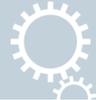

The geometry optimizations and electronic structure calculations are based on the density functional theory (DFT) implemented in the Dmol[3] package with the all-electron double numerical atomic orbital plus polarization (DNP) basis set[42]. The generalized gradient approximation (GGA) with the Perdew-Wang (PW91) exchange-correlation functional[43] is adopted. To account for the dispersion interaction between graphene, a semi-empirical dispersion-correction approach is used (DFT-D)[44]. The dipole correction is also used to eliminate the spurious interaction between the dipole moments of periodic images in the $z$ direction. The Monkhorst-Pack $k$-point mesh[45] is sampled with a separation of about 0.01 Å$^{-1}$ in the Brillouin zone. All the structures are fully relaxed until the maximum force on each atom is less than $10^{-3}$ eV/Å. The component of the energy band is analyzed with resort to additional band structures based on the plane-wave basis set with a cut-off energy of 400 eV and the projector-augmented wave (PAW) pseudopotential implemented in the Vienna ab initio simulation package[46].

Two-probe model is built to simulate the transport of Cu-adsorbed trilayer graphene, and pristine ABC-stacked trilayer graphene is used as electrodes for simplicity. Transport properties are calculated using DFT coupled with the non-equilibrium Green's function (NEGF) formalism implemented in the ATK 11.8 package[47,48]. The single zeta (SZ) basis set is employed. The $k$-points of the electrodes and central region, generated by the Monkhorst-Pack scheme[45], are set to $20 \times 1 \times 50$ and $20 \times 1 \times 1$, respectively. The real-space mesh cutoff is 150 Ry, and the local-density-approximation (LDA) is employed for the exchange–correlation functional. The electrode temperature is set to 300 K.

### Acknowledgements
This work was supported by the National Natural Science Foundation of China (Nos. 11274016, 51072007, 91021017, 11047018, and 60890193), the National Basic Research Program of China (Nos. 2013CB932604 and 2012CB619304), Fundamental Research Funds for the Central Universities, National Foundation for Fostering Talents of Basic Science (No. J1030310/No. J1103205), Program for New Century Excellent Talents in University of MOE of China, and Nebraska Research Initiative (No. 4132050400) and DOE DE-EE003174 in the United States. J. Zheng also acknowledges the financial support from the China Scholarship Council.

### Author contributions
The idea was conceived by J.L. The DFT electronic band calculation was performed by J.M., Z.Z., K.T. and R.Q. The DFT calculation involving the component analysis of the energy band was performed by R.Q., L.W., J.Z., Y.W. and Z.N. The data analyses were performed by J.L., Z.G., J.S. and R.Q. This manuscript was written by R.Q., J.M., Z.Z. and J.L. All authors contributed to the preparation of this manuscript.

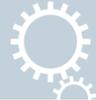

## Additional information

**Supplementary information** accompanies this paper at http://www.nature.com/scientificreports

**Competing financial interests:** The authors declare no competing financial interests.

**License:** This work is licensed under a Creative Commons Attribution-NonCommercial-NoDerivs 3.0 Unported License. To view a copy of this license, visit http://creativecommons.org/licenses/by-nc-nd/3.0/

**How to cite this article:** Quhe, R. *et al.* Tunable band gap in few-layer graphene by surface adsorption. *Sci. Rep.* **3**, 1794; DOI:10.1038/srep01794 (2013).